\documentclass{ws-p10x7}

\begin{document}

\title{
\vspace*{-60pt}
{\normalsize \hfill {\sf UTHEP-435}} \\
\vspace*{-5pt}
{\normalsize \hfill {\sf UTCCP-P-92}} \\
\vspace*{-5pt}
{\normalsize \hfill {\sf September 2000}} \\
\vspace*{30pt}
Dynamical Quark Effects in QCD on the Lattice\newline
--- results from the CP-PACS ---
}

\author{Kazuyuki Kanaya for the CP-PACS Collaboration
}

\address{Institute of Physics, University of Tsukuba,
Tsukuba, Ibaraki  305-8571, Japan\\
E-mail: kanaya@rccp.tsukuba.ac.jp}

\twocolumn[\maketitle\abstract{
Results of a systematic lattice QCD simulation with two degenerate 
flavors of sea quarks, identified as dynamical $u$ and $d$ quarks, 
are presented. 
The simulation was performed on a dedicated parallel computer, 
called CP-PACS, developed at the University of Tsukuba. 
Clear dynamical quark effects are observed in the 
light hadron mass spectrum and in the light quark masses: 
In the light hadron mass spectrum, major parts of the 
discrepancy between quenched QCD and experiment are shown to be removed 
by introducing two flavors of dynamical quarks. 
For the averaged mass of $u$ and $d$ quarks, we find 
$m_{ud}^{\overline{\rm MS}}(2{\rm GeV}) = 3.44^{+0.14}_{-0.22}$ MeV 
using the $\pi$ and $\rho$ meson masses as physical input, and 
for the $s$ quark mass, we obtain 
$m_{s}^{\overline{\rm MS}}(2{\rm GeV}) = 88^{+4}_{-6}$ MeV 
or $90^{+5}_{-11}$ MeV with the $K$ or $\phi$ meson mass as additional 
input. 
These values are about 20--30\% smaller than the previous estimates
in the quenched approximation.
We also discuss the U(1) problem and B meson decay constants.
}]
\footnote{
Talk presented at the XXXth International Conference on High Energy 
Physics (ICHEP 2000), July 27--August 2, 2000, Osaka, Japan. \newline
Current members of the CP-PACS Collaboration are  
A. Ali Khan, S. Aoki, Y. Aoki, G. Boyd, R. Burkhalter, 
S. Ejiri, M. Fukugita, S. Hashimoto, N. Ishizuka, Y. Iwasaki, 
T. Izubuchi, K. Kanaya, T. Kaneko, Y. Kuramashi, T. Manke, 
K. Nagai, J. Noaki, M. Okamoto, M. Okawa, H.P. Shanahan, 
Y. Taniguchi, A. Ukawa, and T. Yoshi\'e.
}

\section{Introduction}

CP-PACS is a dedicated parallel computer designed and developed 
at the University of Tsukuba for simulations in the physics of 
fields\cite{cp-pacs}.
With 2048 node processors interconnected with a three-dimensional 
hypercrossbar network, the CP-PACS achieves a peak performance 
of 614.4 GFLOPS. 
Since 1996, intensive calculations of lattice QCD have been performed 
on the CP-PACS. 
Among others, the first systematic study including both chiral and
continuum extrapolations was attempted for lattice QCD with two 
flavors of dynamical quarks. 
In this paper, we report on the results of these studies, focusing on 
the topics of dynamical quark effects in QCD.

We study lattice QCD\cite{lattice} formulated on a 4-dimensional 
hyper-cubic lattice with a finite lattice spacing $a$.
Continuum physics is defined in the limit of large lattice
volume and vanishing lattice spacing. 
Therefore, 
in order to extract predictions for the real world from the 
simulations on finite lattices, we have to extrapolate 
data obtained on a sufficiently large lattice 
to vanishing lattice spacing ({\it the continuum extrapolation}). 
Furthermore, because the contribution of quarks in the calculation 
is quite computer-time intensive as we decrease the quark mass, 
with the current computers and current algorithms, 
we also have to extrapolate 
to the physical point of light $u$ and $d$ quarks
using data at around the $s$ quark mass region 
({\it the chiral extrapolation}). 
It is important to have good control of the systematic errors due 
to both these extrapolations. 

Because of the huge computational power required, 
majority of calculations have been made in the quenched approximation, 
in which the effects of dynamical quark loops are ignored. 
As the first project on the CP-PACS, we made an extensive simulation 
of quenched QCD\cite{CPPACSquench}. 
The quality of extrapolations and therefore the precision of the 
final hadron spectrum were significantly improved over previous studies. 
From this study, the existence of systematic errors 
due to the quenched approximation was clearly demonstrated 
in the continuum limit.

Therefore, as the next logical step, 
we then performed a series of ``full QCD'' simulations, in which the 
effects of dynamical quarks are taken into account, 
on the CP-PACS\cite{CPPACSfull,CPPACSmq}.
After chiral and continuum extrapolations, 
clear dynamical quark effects are observed in the
light hadron mass spectrum and in the light quark masses.
We also found noticeable effects in B meson decay constants.

In Sec.~\ref{sec:spectrum}, we summarize 
the results for the light hadron spectrum from these studies. 
In Sec.~\ref{sec:Mq}, light quark masses are discussed. 
Sections \ref{sec:U1} and \ref{sec:fB} are devoted to 
the U(1) problem and B meson decay constants, respectively.
Conclusions are given in Sec.~\ref{sec:conclusion}.

\section{Light hadron spectrum}
\label{sec:spectrum}

\begin{figure}
\centerline{
\epsfxsize=6cm \figurebox{}{}{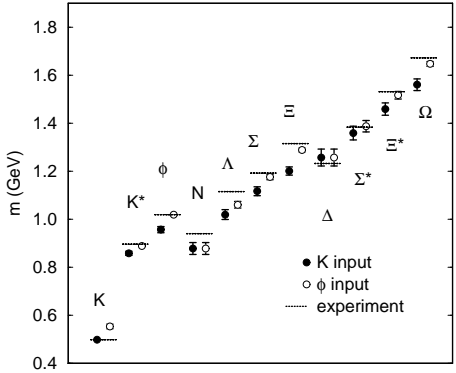}
}
\vspace*{-1mm}
\caption{Quenched light hadron spectrum for ground state mesons 
and baryons in octet and decuplet representations of flavor SU(3).}
\label{fig:qspectrum}
\end{figure}

The precise computation of the hadronic mass spectrum, directly from 
the first principles of QCD, is one of the main goals of lattice QCD. 
This provide us with a direct and non-perturbative test of the validity of 
QCD as the fundamental theory for strong interactions.
 
In Fig.~\ref{fig:qspectrum}, the latest results for the light hadron 
spectrum in the quenched approximation of QCD are 
summarized\cite{CPPACSquench}.
Simulations are made on four lattices $32^3\times56$ to $64^3\times112$ 
with lattice spacings in the range $a\approx 0.1$--0.05 fm.
The spacial lattice size was fixed to be about 3 fm, with which the finite
size effects are estimated to be maximally 0.5\% in the spectrum.
The $u$ and $d$ quarks were treated as degenerate.
On each lattice, five quark masses, corresponding to the 
pseudoscalar-to-vector mass ratio $m_{\rm PS}/m_{\rm V}\approx 0.75$--0.4,
were studied.
The $u,d$ quark mass $m_{ud}$ and the lattice spacing $a$ 
were fixed using
the experimental values for $m_\pi$ and $m_\rho$ as inputs, while the
$s$ quark mass was fixed either by $m_K$ ($K$-input) or $m_\phi$ 
($\phi$-input).
Errors in Fig.~\ref{fig:qspectrum} include statistical as well as systematic 
errors from chiral and continuum extrapolations, but do not include 
the errors from the quenched approximation.

From Fig.~\ref{fig:qspectrum}, we see that, although the global pattern 
of the experimental spectrum is correctly reproduced, there remain systematic 
discrepancies of up to about 10\% (7 standard deviations). 
The resulting spectrum is different depending on the choice of input for 
$s$ quark mass; the $K$-input or the $\phi$-input.
These discrepancies and ambiguities are due to the quenched 
approximation.

\begin{figure}
\centerline{
\epsfxsize=6cm \figurebox{}{}{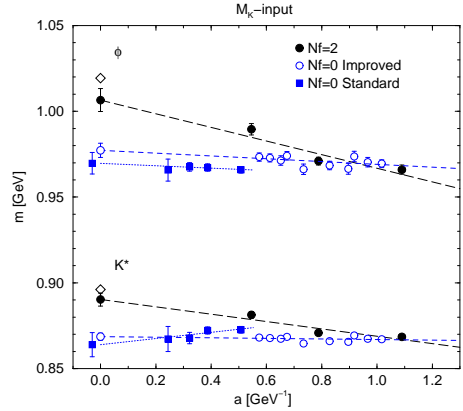}
}
\vspace*{-2mm}
\caption{Continuum extrapolation of vector meson masses $m_{\phi}$ and
$m_{K^*}$ in $N_f=2$ and $N_f=0$ (quenched) QCD, 
using the $K$ meson mass as input.}
\label{fig:fspectrum}
\end{figure}

Because this limitation of the quenched approximation was made clear, 
the next logical step is to perform a ``full QCD'' calculation removing
the quenched approximation. 
As the first step towards the realistic QCD, we performed a series 
of QCD simulations with two degenerate flavors of sea quarks, 
identified as dynamical $u$ and $d$ quarks, while the $s$ quark
is treated in the quenched approximation 
($N_f=2$ QCD)\cite{CPPACSfull,CPPACSmq}.

\begin{table}[b]
\caption{Simulation parameters for $N_f=2$ QCD on the CP-PACS.
$L_s$ is the spacial size of the lattice.
On each lattice, four sea quark masses in the range 
$m_{\rm PS}/m_{\rm V}\approx 0.8$--0.6 were simulated.
For each sea quark mass, we studied hadrons using five valence quarks 
in the range $m_{\rm PS}/m_{\rm V}\approx 0.8$--0.5.
}
\centerline{
\begin{tabular}{cccc}
lattice &  $a$ (fm)  & $L_s$ (fm) & $N_{\rm trajectory}$ \\
\hline
$12^3\times24$ & 0.215(2) & 2.58(3) & 5000--7000 \\
$16^3\times32$ & 0.153(2) & 2.48(3) & 5000--7000 \\
$24^3\times48$ & 0.108(2) & 2.58(3) & 4000 \\
\hline
\end{tabular}
}
\label{tab:fqcdparam}
\end{table}

A key ingredient in avoiding a rapid increase of the computer time is 
the improvement of the lattice theory, with which lattice artifacts 
are reduced on computationally less intensive coarse lattices.
We adopted the combination of an RG-improved 
gauge action and a ``clover''-type improved Wilson quark action,
and carried out the first systematic investigation of full QCD
to perform both continuum and chiral extrapolations. 
Our preparatory full QCD study\cite{compara} shows that this action 
leads already to small lattice artifacts at $a \sim 0.2$ fm.
Therefore, we have chosen the simulation parameters as summarized in 
Table~\ref{tab:fqcdparam}. 
The spacial lattice size was fixed to be about 2.5 fm for all lattices.

Recently, we have doubled the statistics on the finest lattice
at $a \approx 0.1 $fm. 
All results, except for the B meson decay constatnts, presented in 
this paper are based on this full statistics.

Figure~\ref{fig:fspectrum} shows the lattice spacing dependence
of $K^*$ and $\phi$ meson masses from $N_f=2$ QCD, compared with 
the results of quenched calculations. 
For the quenched masses, two different data sets are shown:
Those denoted as ``$N_f=0$ Standard'' are the results of the quenched 
simulation, mentioned before, using the standard lattice 
action\cite{CPPACSquench}.
Because the action used in the full QCD calculation is different 
from the original quenched calculation, 
we carried out an additional quenched simulation using the same 
improved action as for the full QCD runs. The results are denoted as
``$N_f=0$ Improved'' in the figure. 

Our data for hadron spectrum confirms the expectation that 
both quenched calculations must lead to universal values 
in the continuum limit. 
The quenched results, however, show discrepancies from the 
experimental values, as discussed before.
On the other hand, when we introduce two flavors of dynamical quarks, 
the discrepancies are much reduced. 
This means also that the ambiguities from the choice of input for
$s$ quark mass are much reduced in $N_f=2$ QCD.
The remaining small difference might be caused by the quenching 
of the $s$ quark. 

\section{Light quark masses}
\label{sec:Mq}

\begin{figure}[t]
\vspace*{-3mm}
\centerline{
\epsfxsize=7cm \figurebox{}{}{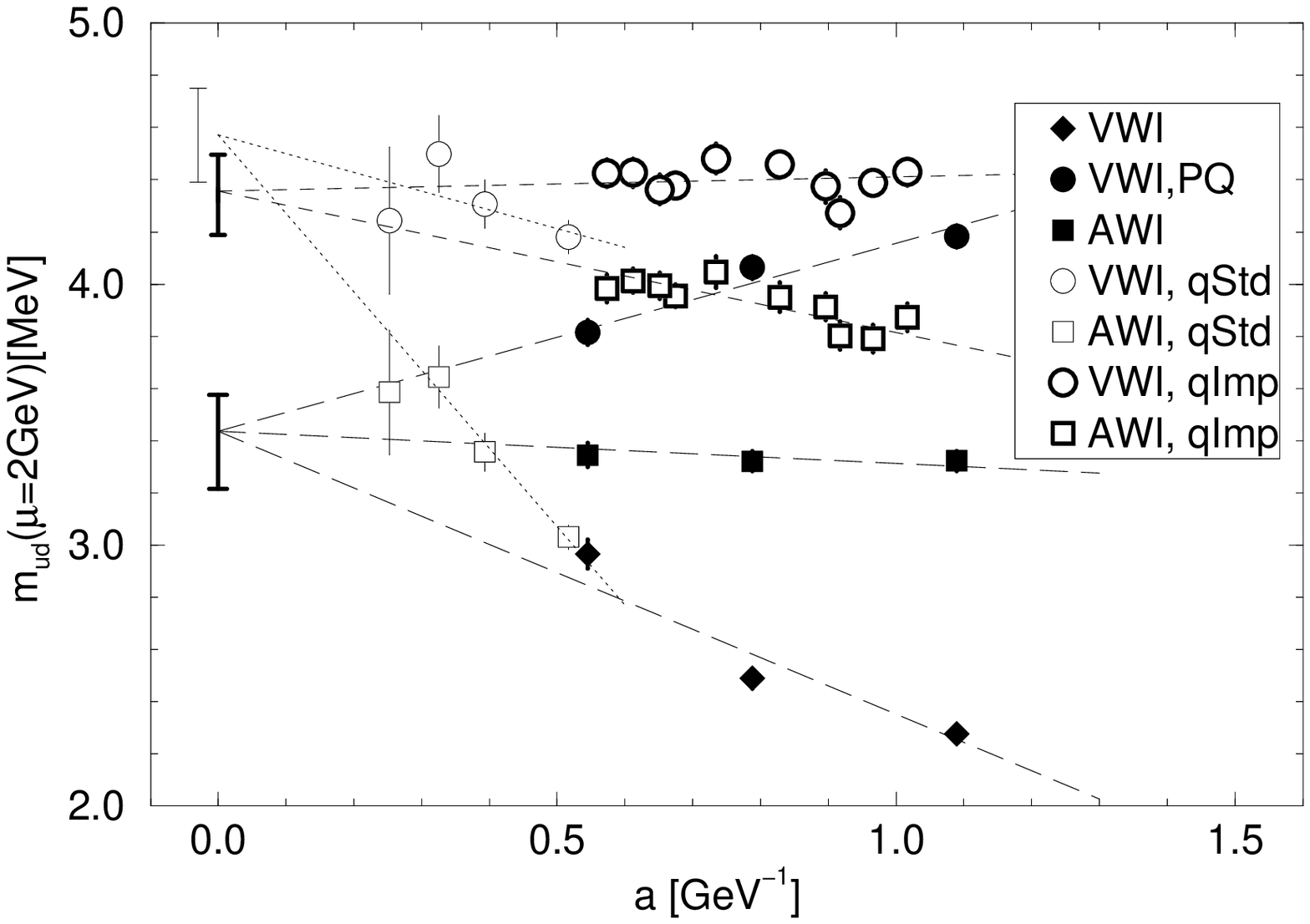}
}
\vspace*{-6mm}
\centerline{
\epsfxsize=7cm \figurebox{}{}{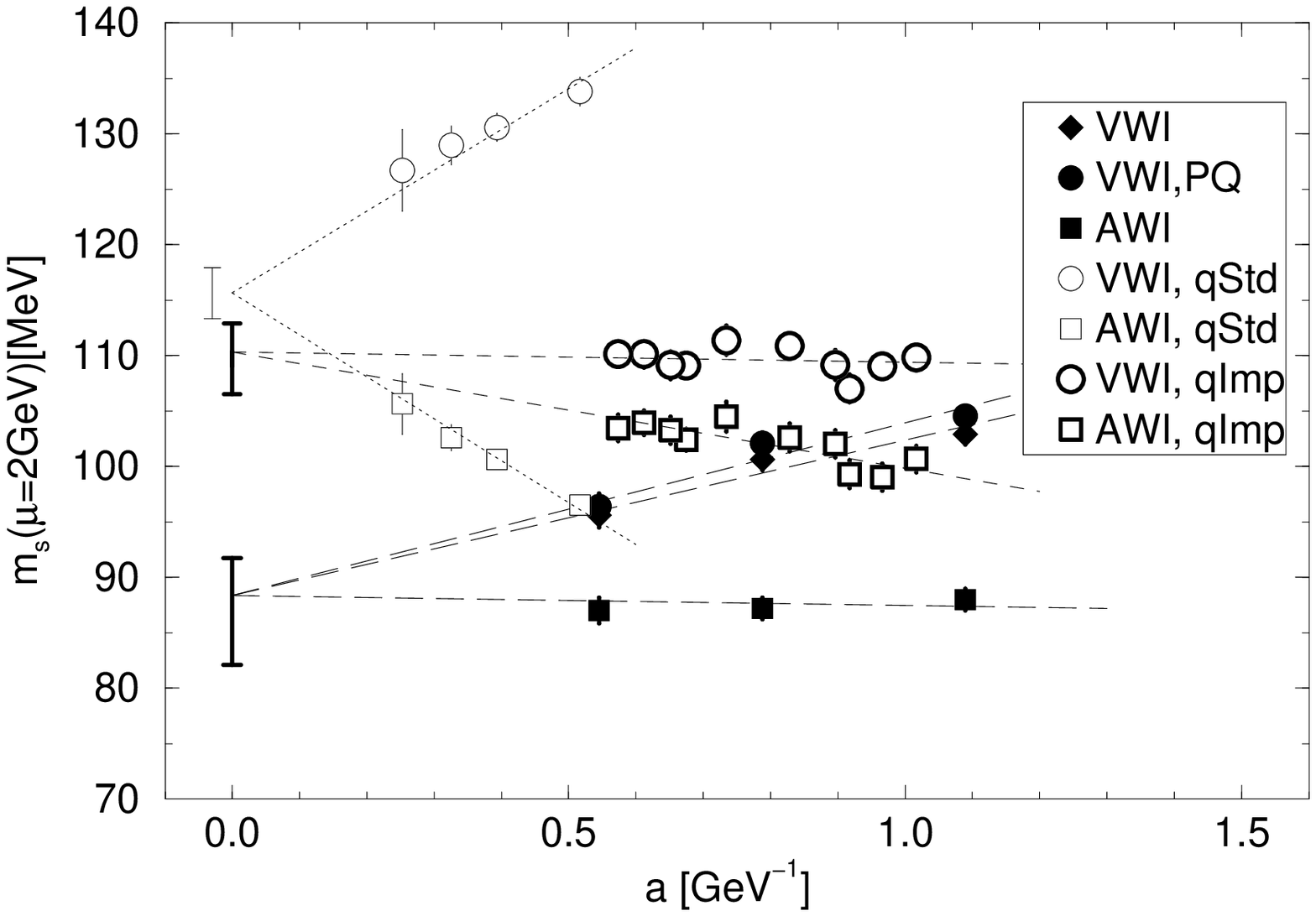} 
}
\vspace*{-5mm}
\caption{Continuum extrapolation of the average $u$ and $d$ quark mass
$m_{ud}$ and the $s$ quark mass $m_s$ in the $\overline{\rm MS}$
scheme at 2 GeV. $m_s$ is from the $K$-input.
Filled symbols are for $N_f=2$ QCD. Quenched results with the standard action
(qStd) and the improved action (qImp) are shown with thin and thick 
open symbols, respectively.}
\label{fig:ms}
\end{figure}

\begin{table*}[t]
\caption{Light quark masses in the $\overline{\rm MS}$
scheme at 2 GeV.}
\centerline{
\begin{tabular}{lccc}
    &  $m_{ud}$ (MeV)  & $m_s$ (MeV) ($K$-input) & 
       $m_s$ (MeV) ($\phi$-input)  \\
\hline
$N_f=0$ standard & 4.57$\pm0.18$          & 116$\pm3$       & 144$\pm6$ \\
$N_f=0$ improved & 4.36$^{+0.14}_{-0.17}$ & 110$^{+3}_{-4}$ & 132$^{+4}_{-6}$ \\
$N_f=2$          & 3.44$^{+0.14}_{-0.22}$ &  88$^{+4}_{-6}$ & 90$^{+5}_{-11}$ \\
\hline
\end{tabular}
}
\label{tab:quarkmass}
\end{table*}

Although quark massses are the most fundamental parameters of QCD, 
due to the confinement, it is impossible to measure them directly 
by an experiment. 
They have to be indirectly inferred from hadronic observables using 
a non-perturbative theoretical relation between these hadronic quantities 
and QCD parameters. 
A lattice QCD determination of the hadron spectrum provides us with 
such a theoretical relation directly from the first principles of QCD. 

Fig.~\ref{fig:ms} summarizes the lattice spacing dependence of 
the average $u$ and $d$ quark mass $m_{ud}$ and the $s$ quark mass $m_s$,
in $N_f=2$ full QCD and in quenched QCD\cite{CPPACSmq}.
On the lattice, there exist several alternative definitions for the quark 
mass. 
In the figures, they are denoted as VWI (vector Ward identity quark masses),
AWI (axial-vector Ward identity quark masses), etc. 
See \cite{CPPACSmq} and \cite{CPPACSquench} for details. 
While different definitions of quark masses lead to results that differ 
at finite lattice spacing, they should converge to a universal value 
in the continuum limit.
Results in Fig.~\ref{fig:ms} clearly demonstrate that 
this is actually the case. 

Values for the light quark masses in the continuum limit are summarized 
in Table~\ref{tab:quarkmass}. 
Errors include our estimates for systematic errors 
from chiral and continuum extrapolations and renormalization factors.
First, we note that the two quenched calculations lead to universal values, 
as in the case of the light hadron spectrum. 
However, the quenched value for $m_s$ differs by about 20\% between 
$K$-input and $\phi$-input. 
We find that 
this discrepancy between the inputs disappears within an error of 10\% 
by the inclusion of two flavors of sea quarks.

The most interesting point is that the values predicted through $N_f=2$ 
QCD are 20--30\% smaller than those in the quenched QCD. 
In particular, our $s$ quark mass in $N_f=2$ QCD is about 90 MeV, 
which is significantly smaller than the value $\approx 
150$ MeV often used in hadron phenomenology, 
and almost saturating an estimate of the lower bound from QCD sum rules 
using the positivity of spectral functions\cite{ChPT-SumRules}.
On the other hand, our result for the $u,d$ to $s$ quark mass ratio,
$m_s^{\overline{\rm MS}}/m_{ud}^{\overline{\rm MS}}=26\pm2$, 
is consistent with $24.4\pm1.5$ from one loop chiral perturbation 
theory\cite{ChPT-Quarkmass}.

\section{U(1) problem}
\label{sec:U1}

\begin{figure}[t]
\centerline{
\epsfxsize=6cm \figurebox{}{}{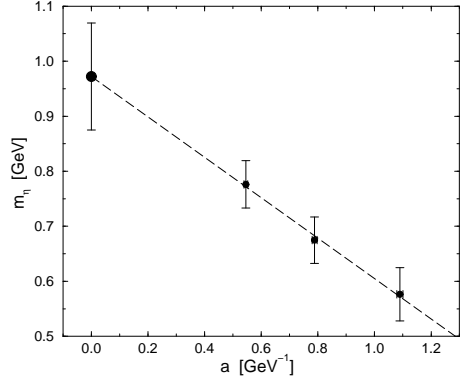}
}
\vspace*{-1mm}
\caption{Continuum extrapolation of the flavor-singlet 
$u\bar{u}+d\bar{d}$ meson mass.}
\label{fig:eta}
\end{figure}

\begin{table}[b]
\caption{Heavy meson decay constants in MeV. 
The lattice scale was fixed by the $\rho$ meson mass. 
Two errors are statistical and systematic. 
For $B_s$ and $D_s$, the $s$ quark mass is fixed from the $K$-input; 
the difference between $K$ and $\phi$-input is found to be smaller than 
the systematic error.
}
\centerline{
\begin{tabular}{llll}
          &          & $N_f=0$ & $N_f=2$ \\
\hline
$f_{B_d}$ & Fermilab & 188$\pm$3$\pm$9  & 208$\pm$10$\pm$11 \\
          & NRQCD    & 191$\pm$5$\pm$11 & 205$\pm$8$\pm$15 \\
\hline
$f_{B_s}$ & Fermilab & 220$\pm$2$\pm$15 & 250$\pm$10$\pm$13 \\
          & NRQCD    & 220$\pm$5$\pm$13 & 242$\pm$8$\pm$17 \\
\hline
$f_{D_d}$ & Fermilab & 218$\pm$2$\pm$15 & 225$\pm$14$\pm$14 \\
$f_{D_s}$ & Fermilab & 250$\pm$1$\pm$18 & 267$\pm$13$\pm$17 \\
\hline
\end{tabular}
}
\label{tab:fB}
\end{table}

The clarification of the mechanism for a large $\eta^\prime$ meson mass 
is an important issue in QCD. 
Propagators of a flavor non-singlet meson consist of a loop of 
a valence quark propagator, while propagators of the flavor singlet 
$\eta^\prime$  meson have an additional contribution 
with two disconnected valence quark loops.
The fact that the $\eta'$ is much heavier than the corresponding 
non-singlet meson $\pi$ means that the two-loop contribution 
should exactly cancel the $\pi$ pole of the one-loop contribution, 
leaving the heavy $\eta'$ pole.
This phenomenon is considered to be related with the anomalous 
violation of the flavor singlet axial U(1) symmetry and 
with the topological structure of gauge field configurations.

The calculation of the two-loop contribution requires a large amount of 
computations on the lattice. 
For this reason only limited results are available. 
In an approximation ignoring the mixing with the $s\bar{s}$ state, 
we studied one and two-loop contributions, and performed, for the 
first time, both chiral and continuum extrapolations\cite{CPPACSfull}.
We obtain $m_{u\bar{u}+d\bar{d}}=972\pm97$~MeV 
for the flavor-singlet $u\bar{u}+d\bar{d}$ meson. 
See Fig.~\ref{fig:eta}. 
In the real world, the $u\bar{u}+d\bar{d}$ state mixes with the 
$s\bar{s}$ state to lead to $\eta$(547) and $\eta'$(958) 
mesons.
We are extending the study to inspect the mixing with the 
$s\bar{s}$ state and the relation to the topological structures.

\section{B mesons on the lattice}
\label{sec:fB}

The decay constant for the $B_q$ meson is defined by
$ 
\langle 0 | \bar{b} \gamma_\mu \gamma_5 q | B_q (p) \rangle
 = i f_{B_q} p_\mu 
$ 
where $q$ denotes either $d$ or $s$ quark.
The non-perturbative determination of $f_{B_q}$, and also 
the bag parameters $B_{B_q}$, 
is quite important for a precise determination of CKM matrix elements. 
Therefore, intensive lattice calculations have been made\cite{latticeheavy}. 

On the lattice, however, the simulation of the heavy $b$ quark is 
not a trivial
extension of light quark simulations, because $m_b \sim 4$ GeV is 
larger than the lattice cutoff $\sim 1$--4 GeV to date. 
Two methods have been developed to simulate heavy quarks on the lattice. 
One is based on a non-relativistic effective theory of QCD 
(NRQCD) defined through an expansion in the inverse heavy quark 
mass\cite{NRQCD}.
Another employs a relativistic action and reinterprets it in terms
of a non-relativistic Hamiltonian (Fermilab method)\cite{Fermilab}.
Because the both methods include an effective treatment of heavy 
quarks, the consistency of the results among them should be checked.

Majority of the lattice studies are done in the quenched approximation. 
On the other hand, a chiral perturbation theory\cite{qChPT} suggests 
sizable corrections from dynamical quarks in the values of $f_{B_q}$.
The first full QCD calculations of $f_{B_q}$ were made 
by the MILC Collaboration\cite{MILC99} using the Fermilab method, 
and by Collins et al.\cite{SGO99} using the NRQCD method.
In these studies, configurations were generated using the 
staggered sea quarks, which is different from the valence light quark 
(the Wilson quark\cite{MILC99} or the clover quark\cite{SGO99}).

Using the CP-PACS computer,
we studied heavy meson decay constants applying a consistent formulation 
for sea and valence light quarks, the clover quark, and
applied both the NRQCD method and the Fermilab method\cite{CPPACSfB}.
Our best estimates of heavy meson decay constants for $N_f=2$ 
and $N_f=0$ (with improved action) are summarized in Table~\ref{tab:fB}.
Because the Fermilab method is applicable also for the $c$ quark, 
we also computed $f_{D_d}$ and $f_{D_s}$ with this method. 
The fact that our $f_{D_s}$ for $N_f=2$ is consistent with the recent 
experimental results, 285$\pm$20$\pm$40 MeV (ALEPH\cite{ALEPH}) and
280$\pm$19$\pm$44 MeV (CLEO\cite{CLEO}), is quite encouraging.

From the table, we see that 
Fermilab and NRQCD methods are consistent with each other. 
We also note that 
$N_f=2$ results for B mesons are about 10--15\% larger than the 
quenched values, while $f_{D_d}$ and $f_{D_s}$ are less 
sensitive to $N_f$. 
Increase of the B meson decay constants affects the determination of
several CKM matrix elements through the $B_q - \overline{B}_q$ mass 
difference $\Delta M_q$. 
Our results for $f_{B_d}$ and $f_{B_s}$ 
are consistent with 
the hypothesis that the Wolfenstein parameter $\rho$ is positive.

\section{Conclusions}
\label{sec:conclusion}

We performed the first systematic study of lattice QCD 
with two flavors of dynamical quarks.
We found that dynamical quark effects are quite important in 
the hadron physics. 
The effect is as large as 20--30\% in the values of light quark mass
and about 10--15\% in B meson decay constants. 
Both of the shifts has significant implications to phenomenological 
studies of the standard model.
It is urgent to evaluate dynamical quark effects in other 
hadronic quantities, such as the bag parameters $B_{B_q}$. 
It is also important to study the effects of dynamical $s$ quark. 
Further intensive studies on the lattice are under way 
to clarify the precise structure of the standard model. 

\section*{Acknowledgments}
The studies presented here were performed by the CP-PACS 
Collaboration.
I thank other members of the Collaboration;
A. Ali Khan, S. Aoki, Y. Aoki, G. Boyd, R. Burkhalter, 
S. Ejiri, M. Fukugita, S. Hashimoto, N. Ishizuka, Y. Iwasaki, 
T. Izubuchi, T. Kaneko, Y. Kuramashi, T. Manke, 
K. Nagai, J. Noaki, M. Okamoto, M. Okawa, H.P. Shanahan, 
Y. Taniguchi, A. Ukawa, and T. Yoshi\'e, for discussions.
This paper is in part supported by 
the Grants-in-Aid of Ministry of Education, Science and Culture 
(No.~10640248) and JSPS Research for Future Program.
 

\end{document}